\documentclass[aps,pre,twocolumn,floatfix]{revtex4}

\usepackage{graphicx}
\usepackage{dcolumn}
\usepackage{bm}

\newcommand{\grad}{\mbox{grad}}

\newcommand{\diffl}[2]{\frac{d #1}{d #2}}

\begin{document}


\title{Sensitivity of rocky planet structures to the equation of state}

\date{June 10, 2009 -- LLNL-TR-414058}

\author{Damian C. Swift}
\affiliation{%
   PLS-CMMD, Lawrence Livermore National Laboratory,
   7000 East Avenue, Livermore, California 94550, USA
}

\begin{abstract}
Structures were calculated for Mercury, Venus, Earth, the Moon, and Mars,
using a core-mantle model and adjusting the core radius to reproduce the
observed mass and diameter of each body.
Structures were calculated using Fe and basalt equations of state 
of different degrees of sophistication for the core and mantle.
The choice of equation of state had a significant effect on the inferred
structure.
For each structure, the moment of inertia ratio was calculated 
and compared with observed values.
Linear Gr\"uneisen equations of state fitted to limited portions of
shock data reproduced the observed moments of inertia significantly better 
than did more detailed equations of state incorporating phase transitions,
presumably reflecting the actual compositions of the bodies.
The linear Gr\"uneisen equations of state and corresponding structures
seem however to be a reasonable starting point for comparative simulations
of large-scale astrophysical impacts.
\end{abstract}


\maketitle

\section{Introduction}
The pressure-temperature-compression equation of state (EOS) 
of condensed matter is vital in understanding planetary structures,
and the response of astrophysical bodies to impacts.
Even for Earth, it is necessary to infer the internal structure of the
planet from limited data.  The situation is much more difficult
for other bodies, where seismic data is at best extremely limited and
more usually absent.

For planetary structures, we are usually interested in pressures from
zero to a few hundred gigapascals.
In this range, EOS were originally deduced empirically from
shock wave measurements \cite{shock}.
Shock experiments typically explore states along the principal Hugoniot
of a material, which may be significantly hotter than the corresponding
compression or pressure occurring in a self-gravitating body,
and corrections are made to estimate the pressure away from the Hugoniot
\cite{shock}.
More recently, mechanical presses such as diamond anvil cells
have provided a source for isothermal compression in this regime,
although the pressure calibration is made with respect to shock
measurements on reference materials \cite{press}.
Samples may also be heated within presses, allowing a wide range of 
states to be explored.
In parallel with the evolution of experimental methods, theoretical techniques
have been developed to predict the EOS from first principles,
typically with an {\it a priori} accuracy of a few percent in mass density,
and better if adjusted to reproduce the STP state \cite{Swift_SiEOS_2001}.
Theoretical EOS are particularly useful as thermodynamically complete states
can be calculated for arbitrary loading conditions that may be
experimentally difficult either to induce or to measure.
Theoretical EOS can also be constructed fairly readily for different
material compositions \cite{Swift_NiAlEOS_2007}.
The prediction of phase diagrams is however typically less certain than the
EOS for a single, known phase.

The strength, failure, and plastic flow behaviors of materials
are also important in planetary physics.
The effects of strength must be taken into account when deducing
the EOS from dynamic loading measurements, as these are at high strain rates
and non-hydrostatic conditions in contrast to the quasistatic conditions
prevailing within self-gravitating bodies.
Conversely, these properties are important for understanding
dynamical processes including the formation of the body,
the effect of impacts, cooling, volcanism, mantle convection, 
and plate tectonics.
These constitutive properties are very challenging to predict theoretically,
and models rely heavily on experimental measurements.
Recent studies have demonstrated the importance of strength in 
the heating induced by shocks and other dynamic loading \cite{Swift_NRS_2008}.
Strength and flow stress can also change greatly with pressure
and temperature.
Recent advances in the measurement and modeling of plastic flow
under extreme conditions has implications for astrophysical impacts,
such as the interpretation of previous simulation studies using
simpler material models.
An example is the widely-quoted study that a liquid interior
allow impact energy to be transported more efficiently to the 
antipode of an impact than would solid components \cite{Hughes1977}.
This conclusion relies strongly on the assumptions used for
material strength.
If the flow stress increases sufficiently with compression,
solid components can transport compression waves more efficiently
than can liquids.

We are constructing theoretical EOS for planetary materials of
different composition, for use in simulations of astrophysical impacts,
focusing on systematic differences in the shape and location of 
impact-produced structures between the rocky bodies.
As a baseline for trial simulations, 
we report here the construction of structure
models for the rocky planets using existing EOS of different sophistication.
There is significant uncertainty over the composition and temperature
profile of all planetary bodies;
here we use EOS for representative substances rather than for the
current best estimates of composition, because a wider variety of EOS of
different degrees of sophistication exist for our chosen materials.
We optimize a single structure parameter -- the core radius -- 
for each combination of EOS.
This approach is different from the commonly-used solution of the Emden
equation \cite{Lyttleton1965,Lyttleton1969} where the compressibility of
the planetary material is assumed to vary linearly with pressure,
and the parameters in this relation -- i.e., the material properties --
are deduced from the observable properties of the planet.

\section{Isostatic structures}
Consider a spherical body comprising compressible material,
of mass density $\rho(r)$.
Knowing or assuming in addition the distribution of composition
and temperature, $\rho(r)$ implies a pressure distribution $p(r)$.
The condition for isostatic equilibrium is that the
stress induced by pressure variations is balanced by the gravitational
acceleration $g(r)$ and centripetal force:
\begin{equation}
\grad\,p(r) = \omega^2-\rho(r)g(r)
\label{eq:gradp}
\end{equation}
where $\omega$ is the angular speed.

For Newtonian gravitation, by Gauss' theorem, 
$g(r)$ can be expressed in terms of the mass enclosed within a given radius 
$m(r)$:
\begin{equation}
g(r) = \frac{Gm(r)}{r^2}
\label{eq:gr}
\end{equation}
so
\begin{equation}
\diffl{g(r)}r = G\left(\frac 1{r^2}\diffl{m(r)}r-2\frac{m(r)}{r^3}\right).
\label{eq:dgdr}
\end{equation}
$m(r)$ can be calculated simply from the distribution of mass density,
\begin{equation}
m(r) = 4\pi\int_0^{r'}r'^2\rho(r')\,dr'
\label{eq:mr}
\end{equation}
or
\begin{equation}
\diffl{m(r)}r = 4\pi r^2\rho(r).
\label{eq:dmdr}
\end{equation}

In the isostatic equilibrium structure,
Eq.~\ref{eq:dmdr} can be integrated from either the center or the surface
$r=R$ to obtain the complete structure.
In practice, the material state at the center depends on the
material models used, so it is more useful to integrate from the
surface, where $p=0$ and $\rho$ can be worked out simply for the
outermost material.

Usually, the total mass $M$ and outer radius $R$ of the body are known
fairly well from observation of orbits and size respectively,
and the problem is to establish the internal structure.
Starting with the assumption of a layered structure, i.e. an ordering of
composition, and the $p(\rho)$ relation for each layer,
any unknown parameter in the structure can be fitted to give the desired
total mass.

For example, consider a layered structure where the radius of one layer is
unknown.
For a stable configuration, inner layers have a higher density.
Consider integration downward from the surface as above, assuming a trial
value for the unknown radius.
If too large, the total mass $M$ will be reached at some $r > 0$;
if too small, $m(0) > 0$.
These two conditions can be used for a solution of this shooting problem 
by bracketing followed by bisection.

Another remotely observable structural parameter is the moment of inertia
$I$, which can be inferred from the rate of precession of the rotational
pole about the normal to the orbit \cite{Gubbins2007}.
In the absence of seismic data, $I$ is a useful constraint on the structure of
the body.
The moment of inertia for a sphere is
\begin{equation}
I=\frac{8\pi}3\int_0^R\rho(r) r^4\,dr,
\end{equation}
and for self-gravitating bodies is conveniently expressed as the dimensionless
moment of inertia ratio,
$I/MR^2$.

Similar approaches assuming the EOS and deducing isostatic planetary
structures have been reported previously.
Kerley has calculated structures for Jupiter and Saturn,
including rotational flattening,
using the theory of figures and very detailed EOS \cite{Kerley2004}.
Seager et al have predicted mass-radius relationships for solid
exoplanets, using EOS which are less suited to subsequent impact
studies and a different solution
approach involving integration outward from the center of the planet
\cite{Seager2007}.
Grasset et al have similarly predicted mass-radius relationships
for rocky and icy exoplanets, again using EOS which are less suited to
impact studies \cite{Grasset2009}.

\section{Compressibility curves for different equations of state}
Two types of EOS were considered: linear Gr\"uneisen EOS using a
representative part of the principal shock Hugoniot as reference,
and more elaborate EOS constructions incorporating phase transitions
and reproducing detailed features of the shock Hugoniot.
The shock Hugoniot data used to calibrate the Gr\"uneisen EOS
are reliable in the sense of being absolute, mechanical measurements,
but require the additional assumption of a thermal correction model
to calculate states off the principal Hugoniot, as are required here.

There is considerable uncertainty in the composition of planetary
interiors, and detailed EOS are not available for all compositions of interest.
In order to calculate structures for all the rocky planets without
introducing additional degrees of freedom in the individual compositions,
the compositions were taken to be pure Fe in the core, and basalt in the mantle.
At zero pressure, basalt typically comprises $\sim$70\%\ plagioclase
(often andesite) and $\sim$30\%\ olivine, and is thus reasonably representative
of the composition thought to occur through the mantle.
Further subdivision of layers was not considered;
the structure chosen allowed the core radius to be used as a parameter
to ensure the correct total mass for each body.
Similar structure calculations using detailed EOS for other compositions
will be reported separately.

For both Fe and basalt, the shock speed-particle speed relation does not follow
a straight line over the full pressure range of interest.
For basalt, a straight line was chosen that reproduced the principal Hugoniot
at pressures of a few tens of gigapascals \cite{vanThiel1966},
as a representative range for planetary mantles.
Similarly, the fit to Fe shock data was for pressures $\sim$50-500\,GPa
\cite{vanThiel1966}.
The more detailed EOS for basalt was a tabulated Gr\"uneisen form
using three piecewise linear fits to the principal Hugoniot 
\cite{Barnes-Lyon1988}.
The detailed Fe EOS was an equilibrium construction comprising four
solid phases and the fluid, with mixed-phase regions \cite{Kerley1993},
and should be much more accurate away from the principal Hugoniot.

In the absence of accurately-known composition profiles, the use of simple
EOS is desirable as it allows planetary structures to be optimized via the
EOS parameters, though this optimization was not done here.  
It is much more cumbersome to optimize detailed EOS in this
was, particularly when presented in tabular form.

\begin{table}
\begin{center}
\caption{Linear Gr\"uneisen equation of state parameters.}
\label{tab:gruneos}
\begin{tabular}{|l|r|r|}\hline
 & {\bf Fe} & {\bf basalt} \\ \hline
$\rho_0$ (g/cm$^3$) & 7.86 & 2.67 \\
$c_0$ (km/s) & 3.635 & 1.45 \\
$s_1$ & 1.802 & 1.97 \\
$\Gamma$ & 2.604 & 1.5 \\
\hline\end{tabular}
\end{center}
\end{table}

The compression curve used was the principal isentrope.
This is a reasonable approach for large, self-gravitating objects.
One would expect $p(\rho)$ to be close to an isentrope on initial formation,
as accretion proceeds by multiple small impacts followed by some
stratification as dense components sink toward the center of the body.
After formation, the decay of radionuclides (and possibly other processes
such as tidal deformation and ohmic heating) may supply additional heat,
but the surface of the body cools by radiation.
In the adiabatic compression of non-porous condensed matter, 
the thermal contribution to pressure is usually much smaller than the 
repulsion between atoms,
so the difference between the isentrope and the isotherm as extreme limits
is small compared with the other uncertainties in planetary
structure such as the composition.

\begin{figure}
\begin{center}\includegraphics[scale=0.72]{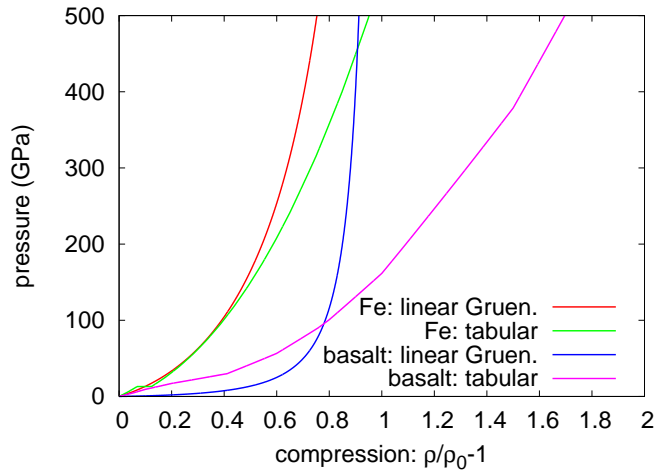}\end{center}
\caption{Comparison between isentropes.}
\label{fig:isencmp}
\end{figure}

Each EOS was used in the form $p(\rho,e)$, 
and integrated numerically starting at STP conditions to obtain $p(\rho)$
along the principal isentrope.
The use of the principal isentrope for each constituent gives a
discontinuity in temperature where different layers meet,
but the effect on pressure is small compared with the uncertainties
in composition and temperature.
It would be straightforward, though less efficient, 
to calculate the isentrope starting at the temperature outside each
layer.
The isentropes for Fe agreed quite closely up to $\sim$120\,GPa;
at higher pressures, the linear Gr\"uneisen EOS was significantly stiffer.
The linear Gr\"uneisen EOS for basalt was much softer than the
tabular EOS at low pressures, and much stiffer at high pressures,
with a crossover at 100\,GPa.
(Fig.~\ref{fig:isencmp}.)

\section{Planetary structures}
The planetary structure equation, Eq.~\ref{eq:dmdr}, was integrated
numerically using the 4th order Runge-Kutta algorithm with a constant
step size in radius.
The effect of rotation was ignored as it depends on latitude;
rotation found to be a negligible effect in a trial equatorial integration.
The core radius was adjusted to give a density profile consistent
with the specified total mass of the body.
The step size was chosen such that core radius was converged numerically 
to 1\,km.
The resulting moment of inertia ratio was converged to 0.1\%, 
which is better than measured values.
The density near the center of the body was very sensitive to small variations
in the core radius or any other parameters, often giving unphysical
variations for the innermost couple of tens of kilometers.
The volume and mass affected were negligible in comparison with the 
overall body.

Using the linear Gr\"uneisen EOS for core and mantle, the numerical
scheme was very robust, and the core radius converged reliably for
very wide ranges of the initial bracket.
Using the tabular EOS, the initial bracket had to be chosen much more
carefully for a physical core radius to be found,
because of the increased compressibility of the core.
compressibility of core.
Converged solutions were however found for all the bodies considered
and compared with observations of the moment of inertia ratio
\cite{Yoder1995} and the usually-quoted values of the core radius
(Table~\ref{tab:struccmp}).
We also summarize pressures calculated at the key locations of
the mantle-core boundary and the center (Table~\ref{tab:pressure})
as an indication of the regimes desirable to explore experimentally,
for each body.

\begin{table}
\begin{center}
\caption{Structure parameters for rocky bodies.}
\label{tab:struccmp}
\begin{tabular}{|l|r|r|r|r|r|r|}\hline
{\bf body} &
   \multicolumn{2}{c|}{\bf reference} &
   \multicolumn{2}{c|}{\bf linear} &
   \multicolumn{2}{c|}{\bf tabular} \\
 &
   \multicolumn{2}{c|}{} &
   \multicolumn{2}{c|}{\bf Gr\"uneisen} &
   \multicolumn{2}{c|}{} \\ \cline{2-7}
 & $r_c$ & $I/MR^2$ & $r_c$ & $I/MR^2$ & $r_c$ & $I/MR^2$  \\
 & (km) & & (km) & & (km) & \\
\hline
Mercury & 1600 & 0.33 & 1740 & 0.3196 & 1802 & 0.3140 \\
Venus & 3000 & 0.33 & 3014 & 0.3348 & 3167 & 0.3146 \\
Earth & 3481 & 0.33 & 3361 & 0.3290 & 3405 & 0.3118 \\
Moon & 350 & 0.393 & 620 & 0.3647 & 752 & 0.3572 \\
Mars & 1700$\pm$500 & 0.366 & 1188 & 0.3621 & 1706 & 0.3316 \\
\hline\end{tabular}
\end{center}
\end{table}

\begin{table}
\begin{center}
\caption{Key pressures in rocky bodies.}
\label{tab:pressure}
\begin{tabular}{|l|r|r|r|r|}\hline
{\bf body} &
   \multicolumn{2}{c|}{\bf linear Gr\"uneisen} &
   \multicolumn{2}{c|}{\bf tabular} \\ \cline{2-5}
 & center & boundary
 & center & boundary \\
 & (GPa) & (GPa) & (GPa) & (GPa) \\ \hline
Mercury & 46.5 & 9.6 & 47.5 & 9.6 \\
Venus & 318 & 121 & 365 & 120 \\
Earth & 390 & 137 & 448 & 143 \\
Moon & 8.8 & 5.0 & 10.0 & 4.4 \\
Mars & 45.2 & 27.2 & 60.0 & 20.7 \\
\hline\end{tabular}
\end{center}
\end{table}

Given that the EOS used were not for the most likely individual 
core compositions that have been postulated (Fe-Ni, or Fe-S for Mars),
and the planetary structures were simplified, it is interesting to
note that the core radius inferred for Earth -- the only body for
which it can be estimated using multiple techniques including seismology --
is too small by only 2\%\ for the tabular EOS and 3\%\ for the linear
Gr\"uneisen.
The core radius for the other bodies did not agree very well with the 
nominal values in the literature, with the exception of Venus,
but the difference was commensurate with the uncertainty in radius.
The exceptions were Venus, where the agreement was good at 0.5-2\%\ 
(the linear Gr\"uneisen EOS being the better), and the Moon,
where the present calculations were larger by a factor $\sim$2.

For the linear Gr\"uneisen EOS, 
the moment of inertia ratios were within 3\%\ except for the Moon,
for which it was too small by 7\%.
For the tabular EOS, the ratio was systematically smaller:
4-5\%\ except for the Moon and Mars (9 and 10\%\ respectively).
Too small a ratio implies that the body is stiffer than the EOS used in
the calculation, in the sense that $p$ would be greater for a given $\rho$.
Such an increase in stiffness could reflect a difference in composition
rather than too low a stiffness for the material used.
Thus the discrepancy for Mars could be attributed to a significant 
proportion of S, reducing the mass density for a given compression.

The linear Gr\"uneisen EOS gives pressures at the center of the Earth
that are within the range of values in the literature (360-390\,GPa).

We next compare in more detail the structures predicted for each body by the
different EOS.

\clearpage
\subsection{Mercury}
\begin{figure}[hb]
\begin{center}\includegraphics[scale=0.72]{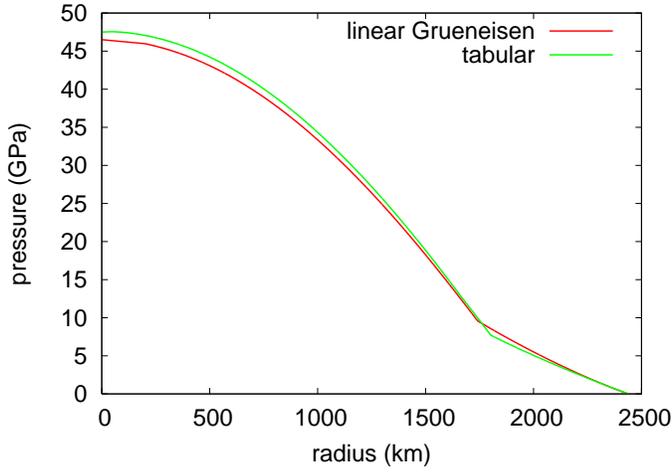}\end{center}
\caption{Radial pressure distribution in Mercury calculated using
   different equations of state.}
\label{fig:mercury_p}
\end{figure}

\begin{figure}[hb]
\begin{center}\includegraphics[scale=0.72]{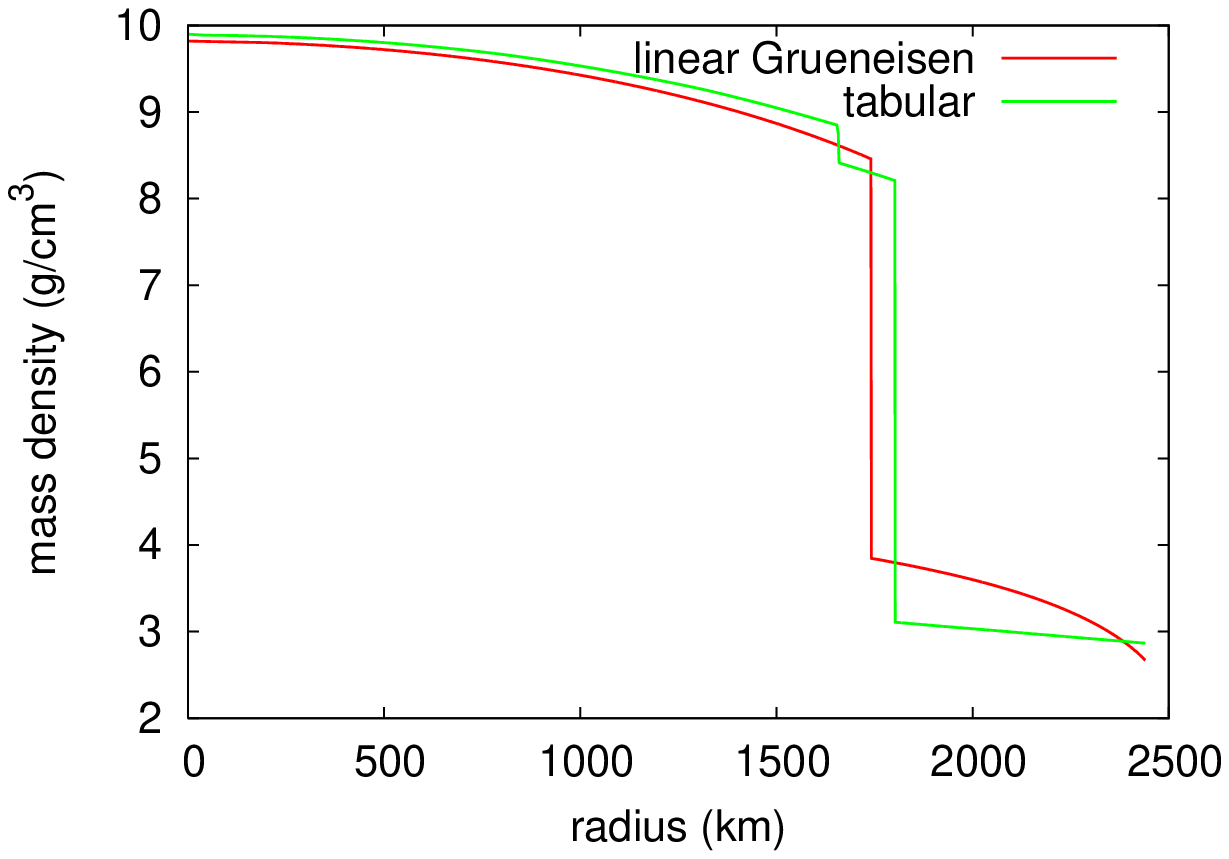}\end{center}
\caption{Radial density distribution in Mercury calculated using
   different equations of state.}
\label{fig:mercury_rho}
\end{figure}

\begin{figure}[hb]
\begin{center}\includegraphics[scale=0.72]{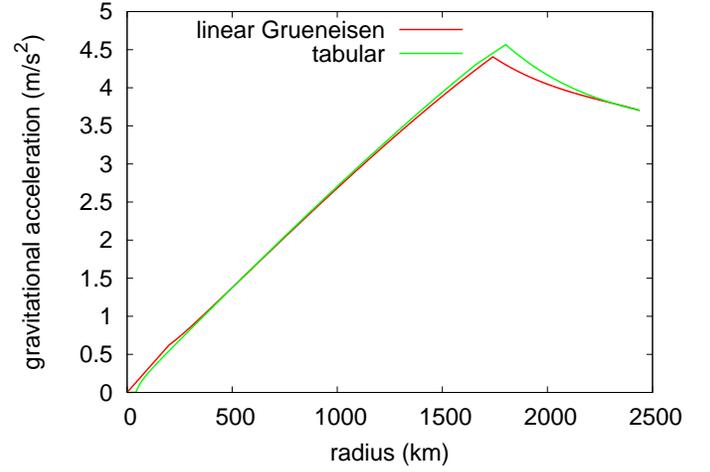}\end{center}
\caption{Radial gravity distribution in Mercury calculated using
   different equations of state.}
\label{fig:mercury_g}
\end{figure}

For Mercury, the EOS gave very similar pressure distributions.
The core radius from each EOS was similar.
The density distribution differed somewhat, particularly in the mantle.
The pressure to induce the $\alpha$-$\epsilon$ phase change occurred just 
within the core, and showed up clearly in the density distribution.
The gravitational acceleration from each EOS was similar.
At the level of detail considered here, the EOS were equivalent.
(Figs~\ref{fig:mercury_p} to \ref{fig:mercury_g}.)

\clearpage
\subsection{Venus}
\begin{figure}[hb]
\begin{center}\includegraphics[scale=0.72]{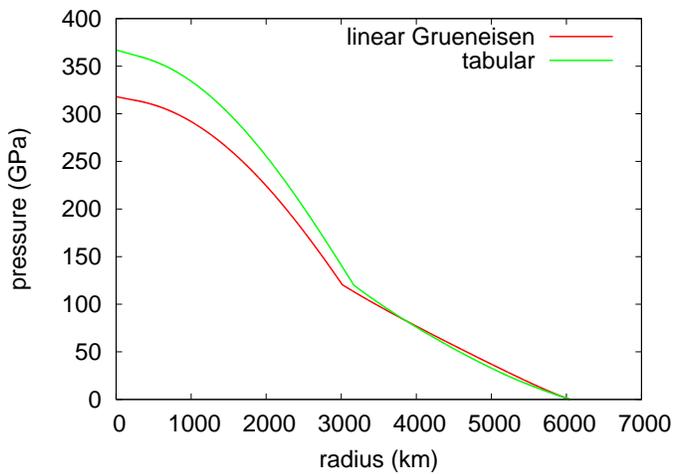}\end{center}
\caption{Radial pressure distribution in Venus calculated using
   different equations of state.}
\label{fig:venus_p}
\end{figure}

\begin{figure}[hb]
\begin{center}\includegraphics[scale=0.72]{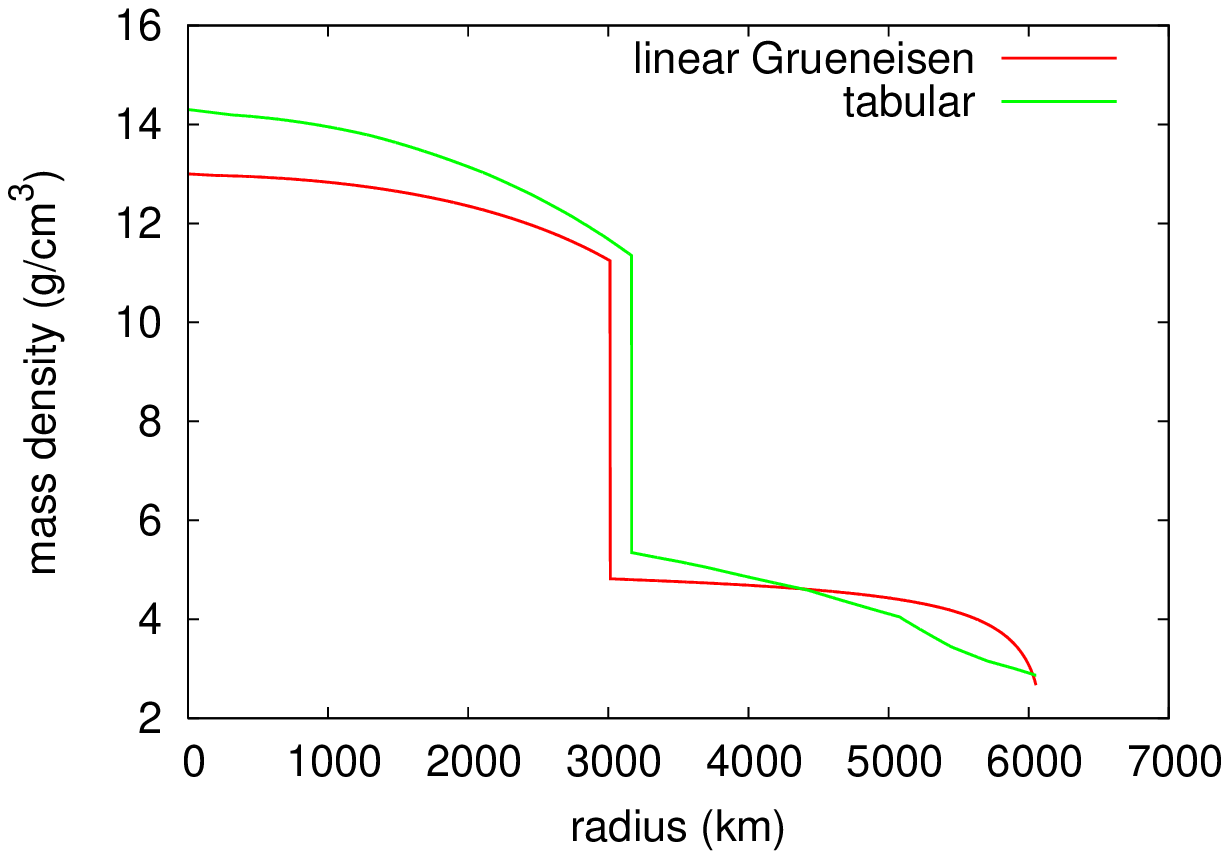}\end{center}
\caption{Radial density distribution in Venus calculated using
   different equations of state.}
\label{fig:venus_rho}
\end{figure}

\begin{figure}[hb]
\begin{center}\includegraphics[scale=0.72]{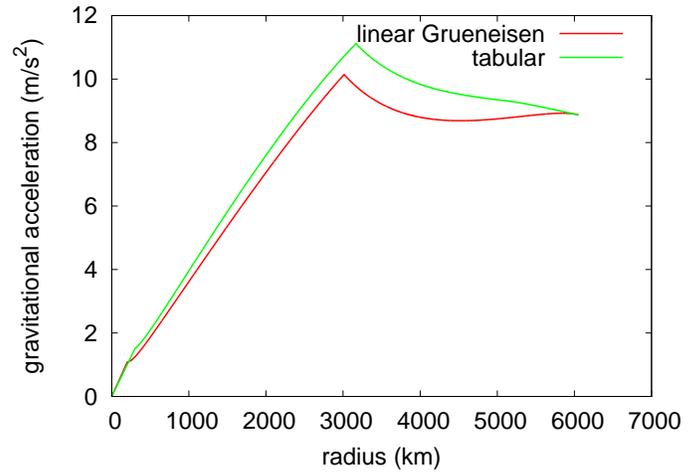}\end{center}
\caption{Radial gravity distribution in Venus calculated using
   different equations of state.}
\label{fig:venus_g}
\end{figure}

For Venus, the EOS gave very similar pressure distributions
in the mantle, and deviated monotonically and significantly in the core.
The core radius from each EOS was similar.
The density distribution differed somewhat.
The core pressure for both EOS was well above the $\alpha$-$\epsilon$
phase transition.
The gravitational acceleration from each EOS was similar within the core,
but differed somewhat in the mantle.
At the level of detail considered here, the EOS were not equivalent
but conversely not hugely different.
(Figs~\ref{fig:venus_p} to \ref{fig:venus_g}.)

\clearpage
\subsection{Earth}
\begin{figure}[hb]
\begin{center}\includegraphics[scale=0.72]{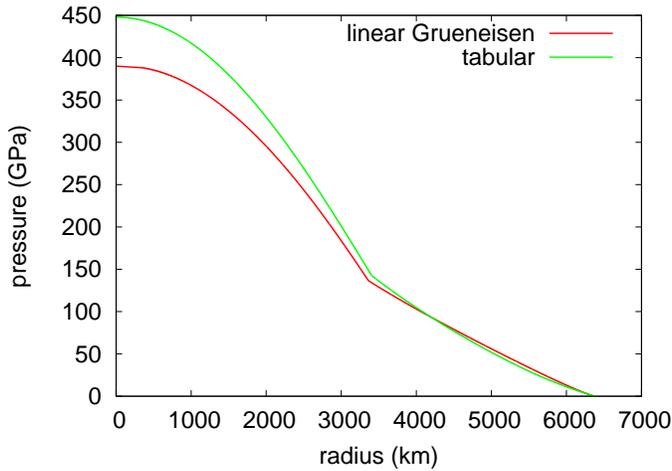}\end{center}
\caption{Radial pressure distribution in Earth calculated using
   different equations of state.}
\label{fig:earth_p}
\end{figure}

\begin{figure}[hb]
\begin{center}\includegraphics[scale=0.72]{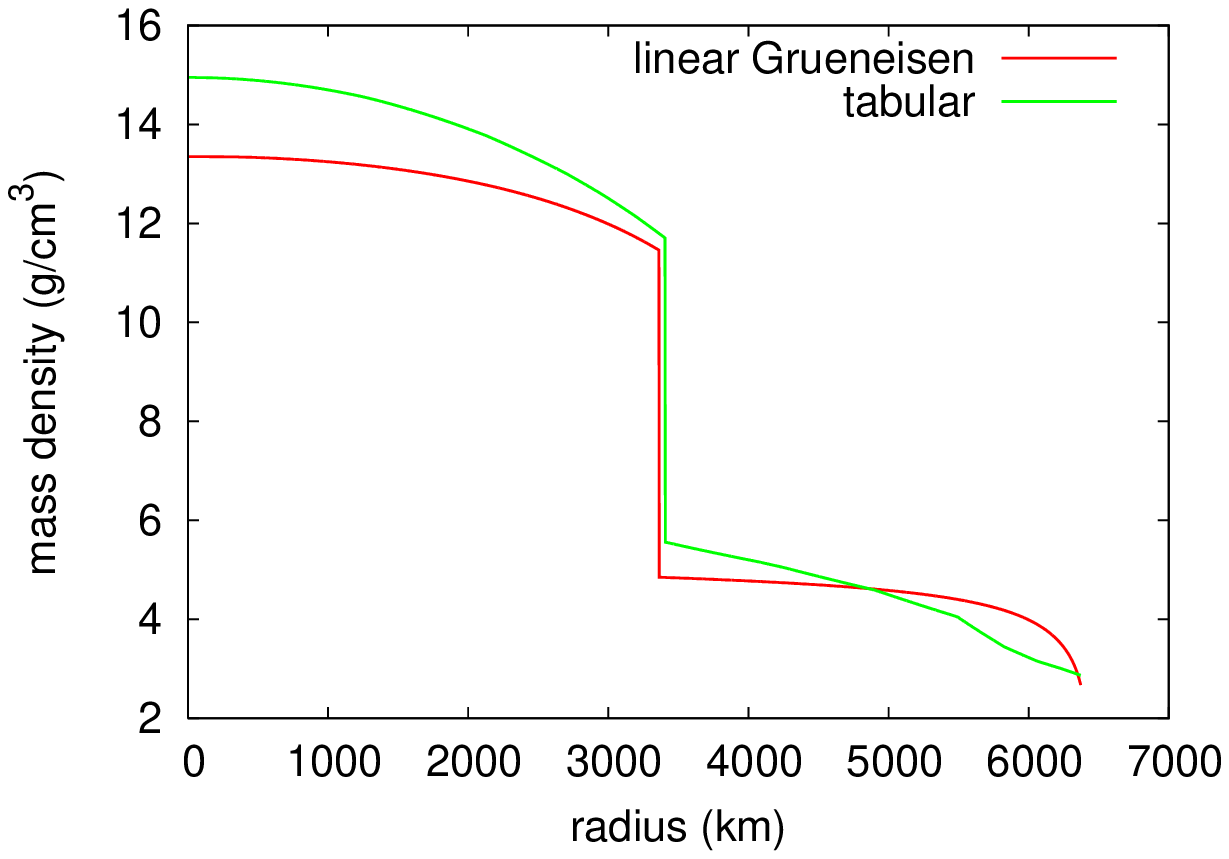}\end{center}
\caption{Radial density distribution in Earth calculated using
   different equations of state.}
\label{fig:earth_rho}
\end{figure}

\begin{figure}[hb]
\begin{center}\includegraphics[scale=0.72]{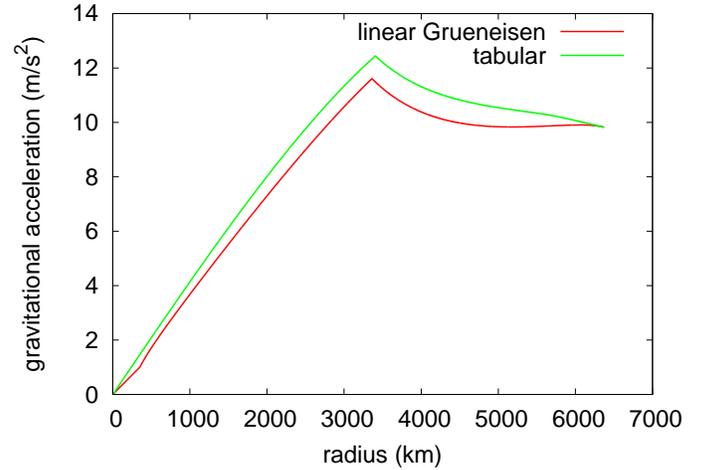}\end{center}
\caption{Radial gravity distribution in Earth calculated using
   different equations of state.}
\label{fig:earth_g}
\end{figure}

The results for Earth were very similar to those for Venus,
as one might expect for bodies of similar mass and volume.
The EOS gave similar pressure distributions
in the mantle, and deviated monotonically and significantly in the core.
The core radius from each EOS was very similar.
The density distribution differed somewhat.
The core pressure for both EOS was well above the $\alpha$-$\epsilon$
phase transition.
The gravitational acceleration from each EOS differed somewhat over the
whole profile (except for the ends, which are constrained to be equal),
although the shapes were very similar.
At the level of detail considered here, the EOS were not equivalent
but not hugely different.
(Figs~\ref{fig:earth_p} to \ref{fig:earth_g}.)

\clearpage
\subsection{Moon}
\begin{figure}[hb]
\begin{center}\includegraphics[scale=0.72]{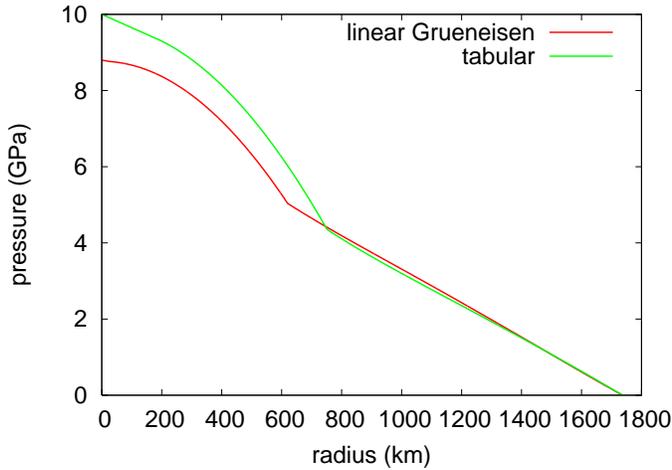}\end{center}
\caption{Radial pressure distribution in the Moon calculated using
   different equations of state.}
\label{fig:moon_p}
\end{figure}

\begin{figure}[hb]
\begin{center}\includegraphics[scale=0.72]{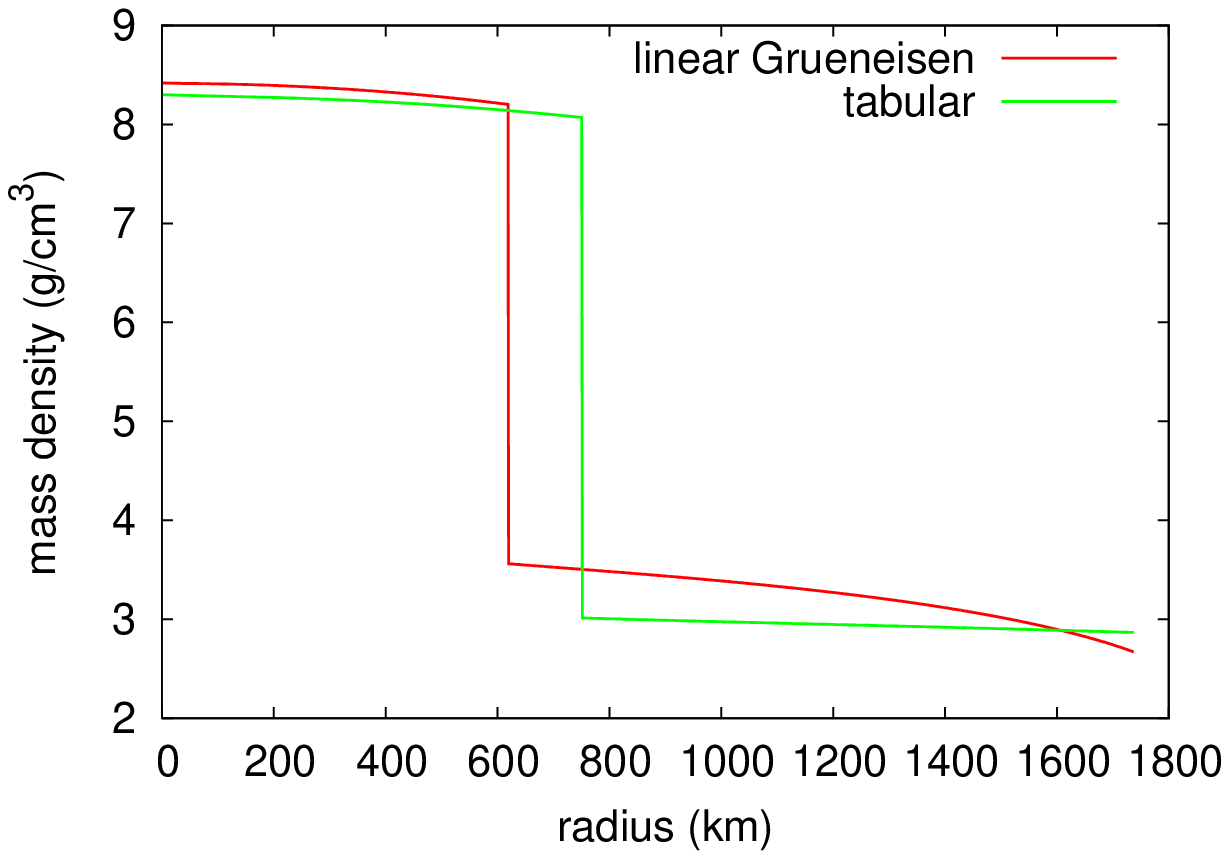}\end{center}
\caption{Radial density distribution in the Moon calculated using
   different equations of state.}
\label{fig:moon_rho}
\end{figure}

\begin{figure}[hb]
\begin{center}\includegraphics[scale=0.72]{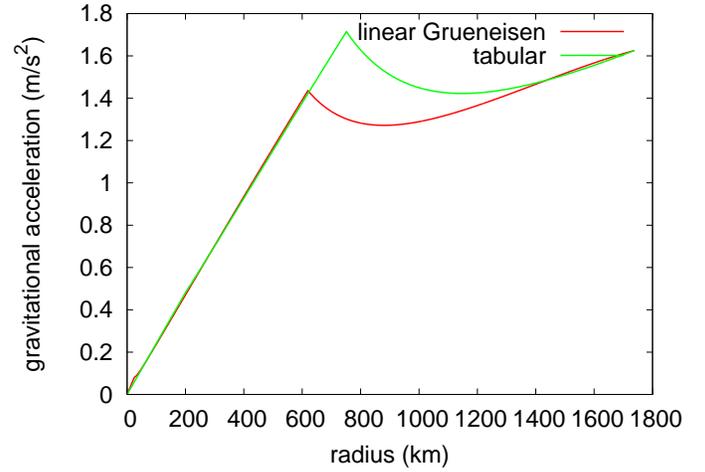}\end{center}
\caption{Radial gravity distribution in the Moon calculated using
   different equations of state.}
\label{fig:moon_g}
\end{figure}

For the Moon, the EOS gave very similar pressure distributions
in the mantle, and deviated monotonically and significantly in the core.
The core radii differed significantly.
The density distributions were very similar, particularly in the core,
except for the difference in core radius.
 differed somewhat, particularly in the mantle.
The core pressures were all below the $\alpha$-$\epsilon$ phase transition.
The gravitational acceleration from each EOS was very similar
in the core, but deviated substantially where the core radii differed
and in the inner mantle.
At the level of detail considered here, the EOS were fairly equivalent.
(Figs~\ref{fig:moon_p} to \ref{fig:moon_g}.)

\clearpage
\subsection{Mars}
\begin{figure}[hb]
\begin{center}\includegraphics[scale=0.72]{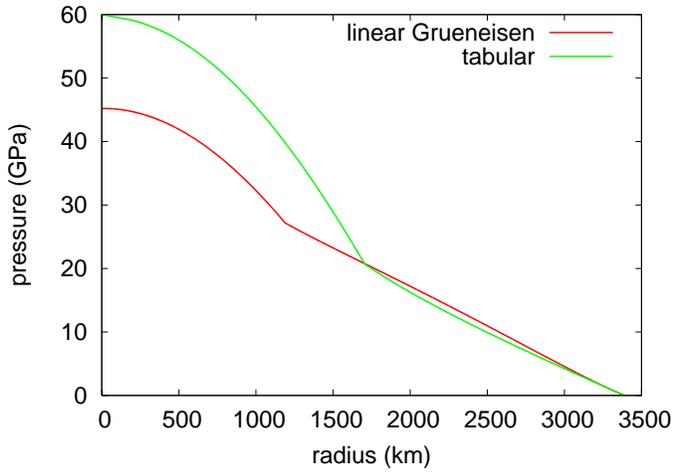}\end{center}
\caption{Radial pressure distribution in Mars calculated using
   different equations of state.}
\label{fig:mars_p}
\end{figure}

\begin{figure}[hb]
\begin{center}\includegraphics[scale=0.72]{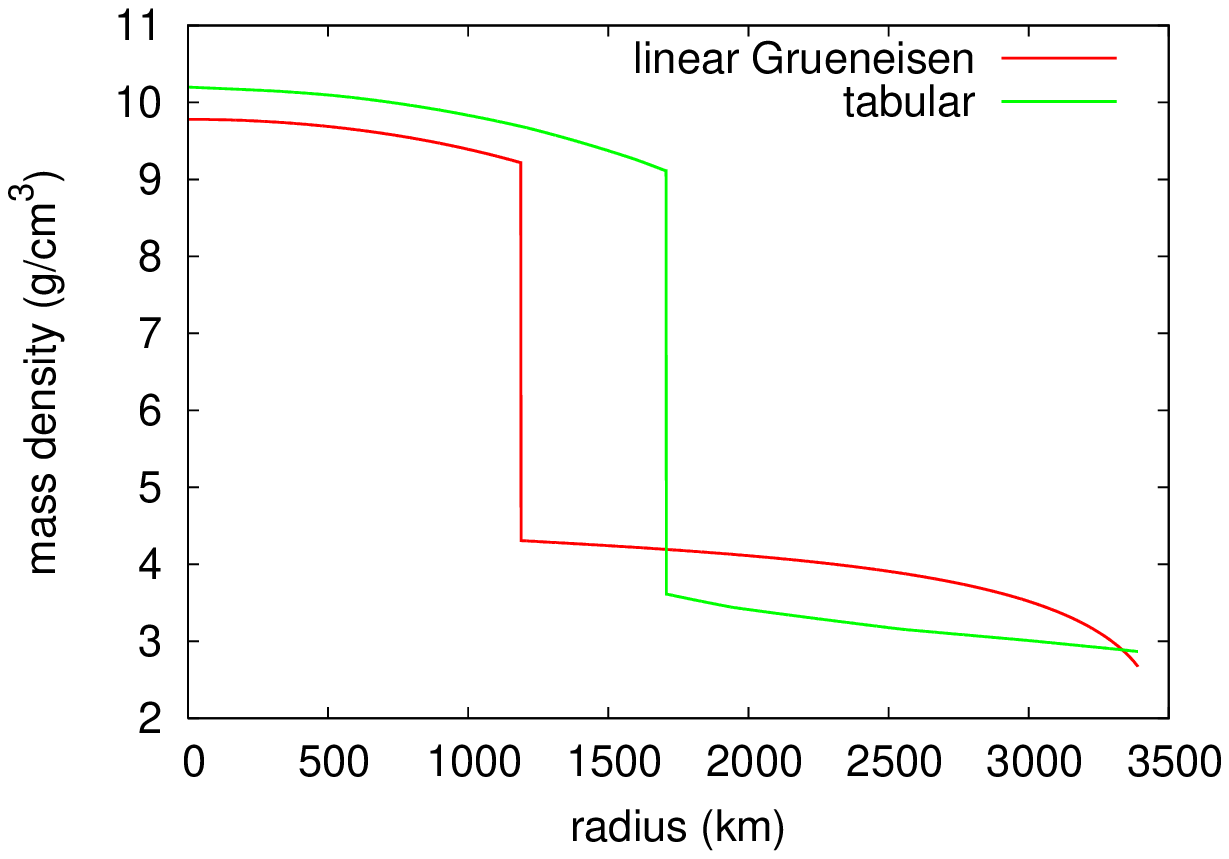}\end{center}
\caption{Radial density distribution in Mars calculated using
   different equations of state.}
\label{fig:mars_rho}
\end{figure}

\begin{figure}[hb]
\begin{center}\includegraphics[scale=0.72]{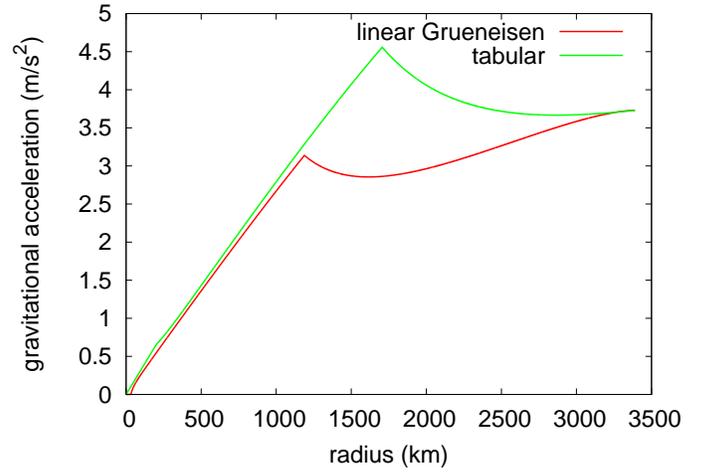}\end{center}
\caption{Radial gravity distribution in Mars calculated using
   different equations of state.}
\label{fig:mars_g}
\end{figure}

For Mars, the EOS gave similar pressure distributions
in the mantle, and deviated greatly in the core.
The core radii differed significantly.
The density distributions were significantly different, but mostly
because of the difference in core radius.
The core pressures were all above the $\alpha$-$\epsilon$ phase transition.
The gravitational acceleration from each EOS was very similar
in the core, but deviated substantially where the core radii differed
and in the mantle.
For simulation purposes, the EOS not equivalent.
(Figs~\ref{fig:mars_p} to \ref{fig:mars_g}.)

\clearpage
\section{Conclusions}
The isostatic equations were used to calculate density distributions
for Mercury, Venus, Earth, the Moon, and Mars, assuming different
EOS for the structural components, and optimizing the core radius
to constrain the overall mass to be correct.
Two pairs of EOS were considered for Fe and basalt,
either linear Gr\"uneisen fits to subsets of published shock data,
or more detailed EOS constructions including phase changes and nonlinearities
in the shock data.
The core radius of Earth, which is by far the best known, was reproduced
reasonably well using both types of EOS.
The other radii, and structures for all the bodies, varied
significantly between the EOS, though to a varying degree.
The structures were used to calculate the moment of inertia ratio for
each body, as an independent test of the accuracy of each structure.

Better overall agreement with the 
observed properties of the rocky bodies considered was obtained
with the linear Gr\"uneisen EOS than the more carefully constructed
tabular EOS.
This does not imply that the linear Gr\"uneisen EOS are more accurate
models of Fe and basalt, but that the properties of the bodies are
represented more closely by the simpler EOS, presumably because the
compositions deviate significantly from Fe and basalt in the core
and mantle respectively.
The simpler EOS do not reproduce the moment of inertia ratios perfectly,
and deviate particularly for the Moon and Mars.

The linear Gr\"uneisen EOS seem to be a reasonable starting point 
for simulations of large-scale impacts on these bodies,
more so than the tabular EOS.
These results also suggest that more detailed EOS for the postulated
compositions of the bodies can be evaluated to some degree by their
improvement over Fe and basalt in calculating the moment of inertia ratios.

\end{document}